\documentclass[prd,a4paper,nofootinbib,floatfix,twocolumn,notitlepage,amssymb,showpacs]{revtex4-1}
\usepackage{amsmath}
\usepackage{graphicx}

\begin{document}

\title{Homoclinic accretion solutions in the Schwarzschild--anti-de Sitter spacetime}
\author{Patryk Mach}
\affiliation{Instytut Fizyki im.~Mariana Smoluchowskiego, Uniwersytet Jagiello\'{n}ski, {\L}ojasiewicza 11, 30-348 Krak\'{o}w, Poland}

\begin{abstract}
The aim of this paper is to clarify the distinction between homoclinic and standard (global) Bondi-type accretion solutions in the Schwarzschild--anti-de Sitter spacetime. The homoclinic solutions have recently been discovered numerically for polytropic equations of state. Here I show that they exist also for certain isothermal (linear) equations of state, and an analytic solution of this type is obtained. It is argued that the existence of such solutions is generic, although for sufficiently relativistic matter models (photon gas, ultra-hard equation of state) there exist global solutions that can be continued to infinity, similarly to standard Michel's solutions in the Schwarzschild spacetime. In contrast to that global solutions should not exist for matter models with a non-vanishing rest-mass component, and this is demonstrated for polytropes. For homoclinic isothermal solutions I derive an upper bound on the mass of the black hole for which stationary transonic accretion is allowed.
\end{abstract}


\maketitle

In a series of 3 papers \cite{karkowski_malec, mach_malec_karkowski, mach_malec} Karkowski, Malec, and I have discussed stationary, Bondi-type accretion flows in Schwarzschild--de Sitter and Schwarzschild--anti-de Sitter spacetimes. We derived analytic solutions for three ``isothermal'' equations of state of the form $p = k e$, where $p$ is the pressure, and $e$ denotes the energy density, with $k = 1/3, 1/2$, and 1. For polytropic equations of state of the form $p = K \rho^\Gamma$, where $\rho$ is the baryonic density and $K$ and $\Gamma$ are constant, we have computed numerical solutions.

While in the sector of the positive cosmological constant $\Lambda$ all these solutions behave in a similar fashion, their behavior differ significantly for negative $\Lambda$. In particular, isothermal transonic solutions with $k = 1/3, 1/2$, and 1 cover the entire space outside the black hole horizon, i.e., they can be continued to arbitrarily large radii (similarly to standard Bondi-type solutions in the Schwarzschild case \cite{bondi, michel}). The situation is different for polytropic solutions, and this behavior is illustrated in Figs.~2 and 3 of \cite{mach_malec_karkowski}. When a polytropic solution is continued outwards, one encounters a finite radius $r$, where the derivative of the radial velocity component $\partial_r u^r$ diverges.

This behavior corresponds to the existence of homoclinic orbits in the phase diagram of the radius $r$ versus the radial component of the velocity $u^r$. Given the equations describing stationary Bondi-type accretion one can construct a dynamical system with the phase portrait consisting of the graphs of different solutions $u^r(r)$ describing the accretion flow \cite{tapas_czerny}. The homoclinic orbit of this dynamical system consists of two transonic solutions --- the solutions that pass through a saddle-type critical (fixed) point, at which the local value of the speed of sound is equal to the modulus of the three velocity of the fluid (the so-called sonic point).

Given the above examples, one could have an impression that those homoclinic solutions are somehow essentially related to the assumed polytropic equation of state (as opposed to the isothermal form). This is not quite true, and the aim of this paper is to clarify these issues.

It is possible to show, and I do that in Sec.~\ref{sec_asymptotic} of this paper, that for isothermal equations of state a power-law asymptotic behavior is admitted only if $k \ge 1/3$. This suggests the existence of homoclinic solutions also for isothermal equations of state with $k < 1/3$, and indeed an analytic example of such a solution can be given for $k = 1/4$ (Sec.~\ref{solk14}).

On the other hand, one can show (Sec.~\ref{sec_poly}) that global (extending to infinity) solutions do not exist for polytropic equations of state, irrespective of the assumed values of the speed of sound. Interestingly, such solutions are not permitted due to the term in the expression for the specific enthalpy that represents the contribution from the rest-mass of gas particles.

The above observations suggest an analogy with the properties of radial time-like and null geodesics in the Schwarzschild--anti-de Sitter spacetime \cite{cruz}. For time-like geodesics the term proportional to $r^2$ in the effective potential forbids the motion of a particle with a finite energy to infinity. In contrast to that, a massless particle can travel to arbitrary large radii.

It seems quite plausible that the occurrence of homoclinic solutions is generic, and it is characteristic for nonrelativistic matter, where the energy density has a nonvanishing contribution from the rest-mass of the gas particles.

Somewhat on the margin of the above considerations I show that for isothermal equations of state with $k < 1/3$ transonic accretion solutions exist only for sufficiently small black holes. The appropriate limit on the mass of the black hole is derived in Sec.\ \ref{criticalpoints}.

Relativistic Bondi-type accretion flows were investigated in many recent papers. For instance, self-gravitating flows were analyzed in \cite{KKMMS, MM2008}, and also in \cite{dok2}. Radiation transfer was included in \cite{lum1, lum2}. Among very recent developments one should also notice \cite{tapas1}. There are also papers dealing with Schwarzschild--anti-de Sitter spacetimes. In \cite{amani} Amani and Farahani discussed phantom accretion onto Schwarzschild--anti-de Sitter black holes, basing on the idea presented in \cite{dok1}. Accretion onto Schwarzschild--anti-de Sitter black holes is also considered in \cite{colvero}.

\section{Notation and equations}

In this paper I assume the gravitational system of units with $c = G  = 1$, and the signature of the metric tensor $(-,+,+,+)$.

The Schwarzschild--anti-de Sitter metric in standard polar coordinates $(t,r,\theta,\phi)$ has the form
\begin{eqnarray}
ds^2 & = & - \left( 1 - \frac{2m}{r} - \frac{\Lambda}{3} r^2 \right) dt^2 + \frac{dr^2}{\left( 1 - \frac{2m}{r} - \frac{\Lambda}{3} r^2 \right)} \nonumber \\
& &  + r^2 \left( d \theta^2 + \sin^2 \theta d \phi^2 \right),
\label{aj}
\end{eqnarray}
where the cosmological constant $\Lambda$ is negative, and $m$ denotes the mass of the black hole. Alternatively, one can work in suitably chosen Eddington--Finkelstein-type coordinates, that are regular at the horizon. An interested reader is referred to \cite{mach_malec_karkowski}.

The horizon of the black hole is located at
\[ r_h = \frac{2}{\sqrt{|\Lambda|}} \sinh \left[ \frac{1}{3} \mathrm{ar \, sinh} \left( 3 m \sqrt{|\Lambda|} \right) \right] \]
(a real positive root of the equation $1 - 2m/r - \Lambda r^2/3 = 0$).

The motion of the fluid is described by the conservation laws
\begin{equation}
\nabla_\mu T^{\mu \nu} = 0, \quad \nabla_\mu (\rho u^\mu) = 0,
\label{wbc}
\end{equation}
where \[ T^{\mu\nu} = (e + p) u^\mu u^\nu + p g^{\mu\nu} \]
is the energy--momentum tensor. Here $e$, $p$, and $\rho$ denote the energy density, the pressure, and the baryonic density, respectively. In what follows we introduce the specific enthalpy $h = (e + p)/\rho$ (note that the specific enthalpy is sometimes defined as $h = (e + p)/\rho - 1$, which is different form the convention used in this paper). The symbol $g^{\mu \nu}$ denotes the components of the metric tensor; $u^\mu$ are the components of the four-velocity of the fluid.

For spherically symmetric and stationary flows $u^\theta = u^\phi = 0$, and all quantities appearing in Eqs.~(\ref{wbc}) are functions of $r$ only. Assuming that the solution is smooth, one can integrate Eqs.\ (\ref{wbc}). This yields
\begin{equation}
\label{www}
h \sqrt{1 - \frac{2m}{r} - \frac{\Lambda}{3}r^2 + (u^r)^2 } = \mathrm{const}, \quad r^2 \rho u^r = \mathrm{const}.
\end{equation}
The above equations constitute a starting point for the discussion of the Bondi-type accretion solutions.
 
In this work I deal with two classes of equations of state. The isothermal equations of state $p = k e$, where $0 < k \le 1$ is a constant \cite{chandra}, and polytropic equations of state $p = K \rho^\Gamma$, where $K > 0$ and $\Gamma > 1$ are constant. In both cases $\rho$, $e$, $h$, and $p$ are related by simple formulas, provided that the flow of the gas is smooth, i.e., there are no shock waves or contact discontinuities. For the isothermal equations of state one has
\begin{equation}
\label{agg}
e = C_1 \rho^{1 + k}, \quad h = (1 + k ) C_1 \rho^k,
\end{equation}
where $C_1$ is a constant. In the case of the polytropic equation of state, it is easy to show that
\begin{equation}
\label{wwz}
 e = C_2 \rho + \frac{K}{\Gamma - 1} \rho^\Gamma, \quad  h = C_2 + \frac{\Gamma}{\Gamma - 1} K \rho^{\Gamma - 1}.
\end{equation}
The choice of the constant $C_2$ is subject to the physical interpretation. In the following I assume standard normalization with $C_2 = 1$. By setting $C_2 = 0$ one can recover Eqs.\ (\ref{agg}).

It is also important to stress that the conservation law $\nabla_\mu (\rho u^\mu) = 0$ in Eqs.~(\ref{wbc}) is, de facto, the definition of $\rho$. This quantity can have different physical interpretation, depending on the context. For the gas of a conserved number of massive particles, the density $\rho$ can be expressed as the number density of the particles times the mean rest-mass of the particle. On the other hand, the dynamics of the photon gas with the equation of state $p = e/3$ is described by equations $\nabla_\mu T^{\mu \nu} = 0$ only. In this case it is also possible to introduce a function $\rho$ satisfying the conservation law $\nabla_\mu (\rho u^\mu) = 0$, provided that the flow of the gas is smooth. Such a function has a natural interpretation of the specific entropy (note that it is not conserved across possible discontinuities in the flow).

By differentiating Eqs.\ (\ref{www}) with respect to $r$ one can obtain the following expression for $du^r/dr$:
\[ \frac{du^r}{dr} = \frac{2 u^r}{r} \frac{c_s^2 \left[ 1 - \frac{2m}{r} - \frac{\Lambda}{3}r^2 + (u^r)^2 \right] - \frac{m}{2r} + \frac{\Lambda}{6} r^2}{(u^r)^2 - c_s^2 \left[ 1 - \frac{2m}{r} - \frac{\Lambda}{3}r^2 + (u^r)^2 \right]}, \]
where $dh/d\rho = h c_s^2 / \rho$, and $c_s^2$ is the speed of sound. One has $c_s^2 = k$ for isothermal equations of state, and $c_s^2 = (\Gamma - 1)(1 - 1/h)$ for polytropes. Formally, this constitutes a ``step backwards'' in the process of finding of solutions, but it gives an insight into their structure. Let $l = l(r)$ be a function of $r$ such that
\begin{eqnarray}
\frac{dr}{dl} & = & r \left\{ (u^r)^2 - c_s^2 \left[ 1 - \frac{2m}{r} - \frac{\Lambda}{3}r^2 + (u^r)^2 \right] \right\} \nonumber \\
& \equiv & f_1(r,u^r).
\label{dyn_a}
\end{eqnarray}
Then
\begin{eqnarray}
\frac{du^r}{dl} & = & 2 u^r \left\{ c_s^2 \left[ 1 - \frac{2m}{r} - \frac{\Lambda}{3}r^2 + (u^r)^2 \right] - \frac{m}{2r} + \frac{\Lambda}{6} r^2 \right\} \nonumber \\
& \equiv & f_2 (r,u^r).
\label{dyn_b}
\end{eqnarray}
Equations (\ref{dyn_a}) and (\ref{dyn_b}) can be treated as a dynamical system, whose phase portrait consists of the graphs of $u^r$ versus $r$, or more precisely, the graphs of $u^r(r)$ belong to the orbits of Eqs.\ (\ref{dyn_a}) and (\ref{dyn_b}). 

The dynamical system defined by Eqs.\ (\ref{dyn_a}) and (\ref{dyn_b}) has critical (fixed) points $(r_\ast, u^r_\ast)$ where $f_1(r_\ast,u^r_\ast) = f_2(r_\ast, u^r_\ast) = 0$, that is
\begin{eqnarray}
\label{crita}
(u^r_\ast)^2 - c_{s \ast}^2 \left[ 1 - \frac{2m}{r_\ast} - \frac{\Lambda}{3}r_\ast^2 + (u^r_\ast)^2 \right] & = & 0, \\
c_{s\ast}^2 \left[ 1 - \frac{2m}{r_\ast} - \frac{\Lambda}{3}r_\ast^2 + (u^r_\ast)^2 \right] - \frac{m}{2r_\ast} + \frac{\Lambda}{6} r_\ast^2 & = & 0.
\label{critb}
\end{eqnarray}
Here, and in what follows, the quantities referring to the critical point will be denoted with an asterisk.

In order to find critical points one has to specify the equation of state. It is, however, clear that
\begin{equation}
\label{trala}
(u_\ast^r)^2 = \frac{m}{2 r_\ast} - \frac{\Lambda}{6} r_\ast^2,
\end{equation}
and also
\[ c_{s\ast}^2 = \frac{\frac{m}{2 r_\ast} - \frac{\Lambda}{6} r_\ast^2}{1 - \frac{3m}{2 r_\ast} - \frac{\Lambda}{2}r_\ast^2}. \]
Thus, one can easily show that at the critical point
\[ c_{s\ast}^2 = (g_{rr} v^r v^r)_\ast = (v_r v^r)_\ast, \]
where $v^r = u^r/u^t$ is the radial component of the three-velocity. Thence it is common to use the term ``sonic point'' instead of ``critical point''. For $\Lambda = 0$, i.e., in the Schwarzschild case, and for standard equations of state, $(r_\ast, u^r_\ast)$ is a saddle point. Thus, there is an accretion solution passing through the sonic point. In the following I will reserve the term ``sonic point'' for a critical saddle point, as opposed to other types of critical points.

The existence and number of critical points will be discussed in Sec.\ \ref{criticalpoints} for the isothermal equations of state. 

\section{Asymptotic behavior of isothermal solutions}
\label{sec_asymptotic}

It is quite illuminating to start with an analysis of the asymptotic behavior of isothermal solutions\footnote{A part of the analysis presented in this chapter was pointed to me by Jerzy Knopik.}. For isothermal equations of state Eqs.~(\ref{www}) yield
\begin{equation}
\label{yyy}
1 - \frac{2m}{r} - \frac{\Lambda}{3}r^2 + (u^r)^2 = A r^{4k} |u^r|^{2k},
\end{equation}
where $A$ is a positive constant. Let us assume an asymptotic expansion of the form $u^r \simeq B r^\alpha$. One gets
\[ 1 - \frac{2m}{r} - \frac{\Lambda}{3}r^2 + B^2 r^{2 \alpha} = A B^{2k} r^{2k(2 + \alpha)}.  \]
There are two ways in which this equation can be satisfied in the leading order for $r \to \infty$. In both cases the term on the right-hand side has to cancel with the leading term at the left-hand side. If the leading order term on the left-hand side is $-\Lambda r^2/3$, then $2k(2 + \alpha) = 2$, and $2 > 2 \alpha$ (the condition that the term  $B^2 r^{2 \alpha}$ is not the leading one). If the leading order term is $B^2 r^{2 \alpha}$, we have $2 k (2 + \alpha) = 2 \alpha$, and $2 \alpha > 2$. These two possibilities lead to $\alpha = (1 - 2k)/k$ and $\alpha = 2k/(1 - k)$, respectively. In either case $k > 1/3$. For $k = 1/3$ the two asymptotics coincide; one has $\alpha = 1$. It can be seen from the examples given in \cite{mach_malec_karkowski} that the two exponents $\alpha$ correspond to the branches of the solution that are asymptotically subsonic and supersonic respectively (note that for $1/3 < k < 1$ we have $(1 - 2k)/k < 2k/(1 - k)$). If $k = 1/3$, the two branches still exist, but with the same asymptotic behavior.

It is clear that for $k < 1/3$ asymptotic behavior of the form $u^r \simeq B r^\alpha$ is not admitted. This suggests that for $k < 1/3$ global solutions do not exist. The following example confirms this expectation.

\section{Solution for $p = e/4$}
\label{solk14}

\begin{figure}
\begin{center}
\includegraphics[width=\columnwidth]{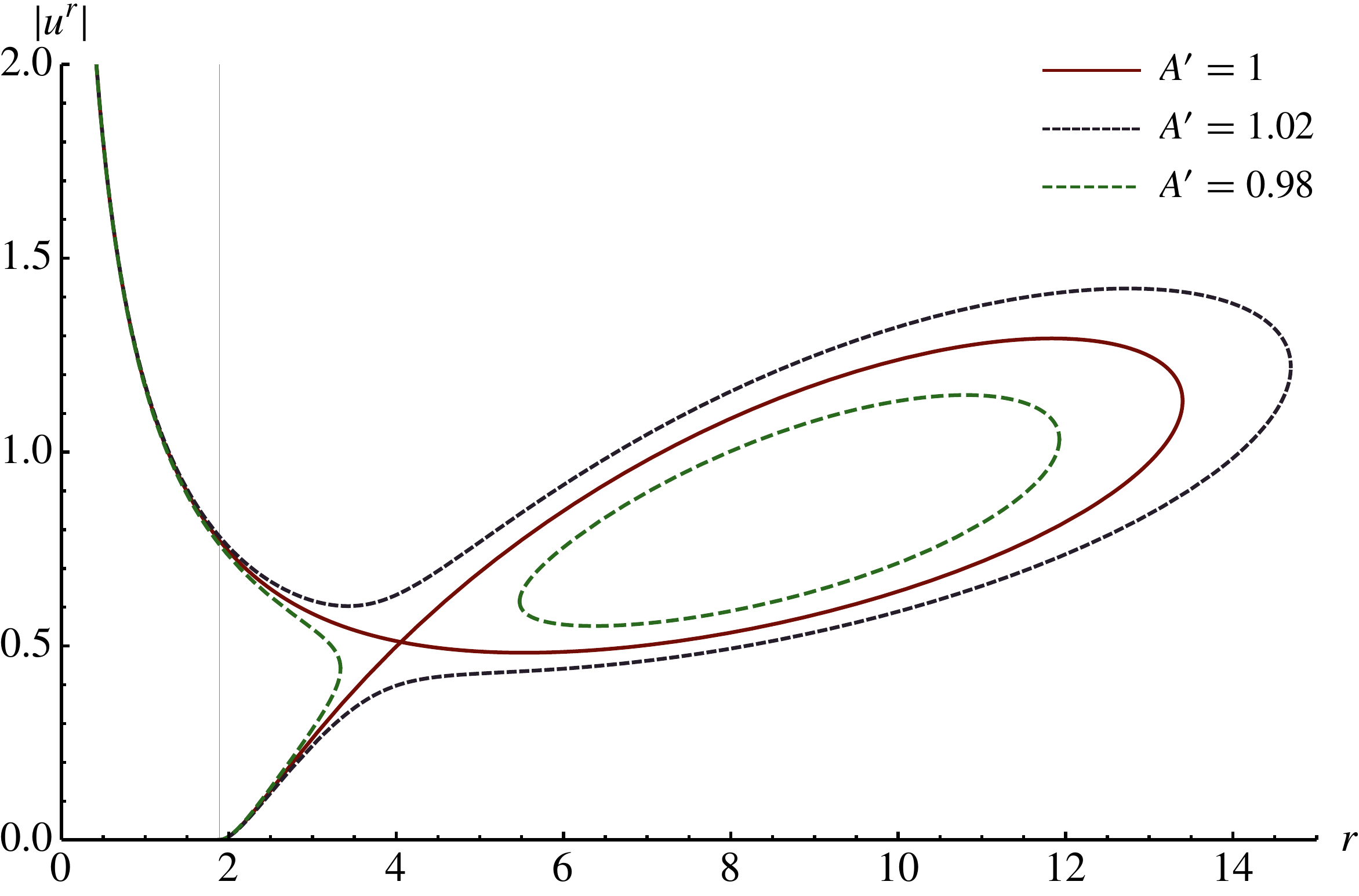}
\end{center}
\caption{Solutions obtained for the equation of state $p = e/4$ and sample metric parameters $m = 1$, $\Lambda = -5/100$. The solid line depicts transonic solutions. The vertical line denotes the position of the horizon of the black hole. The graph shows solutions for $A^\prime = A/A_\ast = 0.98$, 1, and 1.02.}
\end{figure}

An explicit homoclinic solution can be found for the equation of state $p = e/4$. For this equation of state Eq.~(\ref{yyy}) can be written as
\begin{equation}
\label{wwa}
1 - \frac{2m}{r} - \frac{\Lambda}{3}r^2 + (u^r)^2 = A r \sqrt{|u^r|},
\end{equation}
Here I will be mainly interested in transonic solutions. In this case the value of $A$ should be determined by requiring that the solution passes through the sonic point (a saddle critical point of the dynamical system defined by Eqs.\ (\ref{dyn_a}) and (\ref{dyn_b})). The location of the sonic point $r_\ast$ is given by
\[ \frac{m}{2 r_\ast} - \frac{\Lambda}{6} r_\ast^2 = \frac{1}{4} \left( 1 - \frac{3 m}{2 r_\ast} - \frac{\Lambda}{2} r_\ast^2 \right) \]
or
\[ \Lambda r_\ast^3 + 6 r_\ast - 21 m = 0 \]
(cf.\ Sec.\ \ref{criticalpoints}). The above polynomial equation has real positive solutions only for $m^2 \le 32/(441 |\Lambda|)$, i.e., when its discriminant is not positive. This follows from a very simple reasoning; it is presented in Sec.\ \ref{criticalpoints} without restricting to $k = 1/4$ only. For $m^2 \le 32/(441 |\Lambda|)$ the location of the sonic point is given by
\[ r_\ast = \frac{2 \sqrt{2}}{\sqrt{|\Lambda|}} \cos \left[ \frac{\pi}{3} + \frac{1}{3} \mathrm{arc \, cos} \left( \frac{21 m \sqrt{|\Lambda|}}{4 \sqrt{2}} \right) \right]. \]
The square of the radial component of the velocity at the sonic point is given by Eq.\ (\ref{trala}); the value of the constant $A$ appearing in Eq.~(\ref{wwa}) that corresponds to transonic solutions can be expressed as
\begin{equation}
\label{wwb}
A = A_\ast \equiv \frac{4}{r_\ast} \left( \frac{m}{2 r_\ast} - \frac{\Lambda}{6} r_\ast^2 \right)^\frac{3}{4}.
\end{equation}

Equation (\ref{wwa}) is a quartic polynomial equation in $\sqrt{|u^r|}$, and it can be solved exactly. Let
\[ \Delta = \left( \frac{A r}{4} \right)^4 - \frac{1}{3^3} \left( 1 - \frac{2m}{r} - \frac{\Lambda}{3} r^2 \right)^3, \]
where $A = A_\ast$ is given by Eq.~(\ref{wwb}). The domain of the transonic solution is given by the condition $\Delta \ge 0$. Let $R$ denote the largest root of the equation $\Delta = 0$, and let us set
\[ X_\pm = \frac{y}{2} \left( 1 \pm \sqrt{\frac{A r}{\sqrt{2} y^\frac{3}{2}} - 1} \right)^2, \]
\[ y = \sqrt[3]{ \left( \frac{A r}{4} \right)^2 + \sqrt{\Delta}} + \sqrt[3]{ \left( \frac{A r}{4} \right)^2 - \sqrt{\Delta}},  \]
where in the expression for $y$ we choose real roots. Then the two branches of the transonic solution can be written as
\[ |u^r| = \begin{cases} X_+, & 0 < r \leq r_\ast, \\ X_-, & r_\ast \leq r < R, \end{cases} \]
(this branch is subsonic outside $r_\ast$) and
\[ |u^r| = \begin{cases} X_-, & r_h < r \leq r_\ast, \\ X+, & r_\ast \leq r < R, \end{cases} \]
for the branch that is supersonic outside $r_\ast$. For values of the constant $A$ other than that given by Eq.~(\ref{wwb}) the expressions $X_\pm$ give remaining accretion solutions, that are not transonic.

Once the expressions for $u^r$ are found, all other quantities can be trivially computed. The density $\rho$ is given by the second equation in Eqs.~(\ref{www}). The expressions for $e$ and $h$ are obtained from Eqs.\ (\ref{agg}).

Examples of solutions for the equation of state $p = e/4$ are depicted in Fig.\ 1. Here $m = 1$, $\Lambda = -5/100$, and the solutions are plotted for $A^\prime \equiv A/A_\ast = 0.98, 1$, and 1.02, respectively.

\section{Critical points in the isothermal case}
\label{criticalpoints}

The discussion of the asymptotic behavior of isothermal solutions given in Sec.\ \ref{sec_asymptotic}  and the solutions shown in Fig.\ 1 suggest that for isothermal equations of state $p = k e$ with $0 < k < 1/3$ one should expect the existence of two critical points on the phase diagram of $(r,u^r)$, at least for a certain range of parameters. These should be a saddle point belonging to the homoclinic orbit, and a center enclosed by this orbit. I will now argue that this is indeed the case. In addition, a simple reasoning gives a restriction on the mass of the black hole allowing for transonic accretion.

For isothermal equations of state $p = k e$ Eqs.\ (\ref{crita}) and (\ref{critb}) yield 
\[ \frac{m}{2 r_\ast} - \frac{\Lambda}{6} r_\ast^2 = k \left( 1 - \frac{3 m}{2 r_\ast} - \frac{\Lambda}{2} r_\ast^2 \right), \]
which is equivalent to the cubic equation
\[ \frac{\Lambda}{2} \left( \frac{1}{3} - k \right) r_\ast^3 + k r_\ast - \frac{3}{2}m \left( k + \frac{1}{3} \right) = 0. \]
Dividing by $\frac{\Lambda}{2} \left( \frac{1}{3} - k \right)$ one can write the above equation as
\begin{equation}
\label{eqf}
f(r_\ast) \equiv r_\ast^3 + 3 p r_\ast + 2 q = 0,
\end{equation}
where
\[ p = \frac{2k}{\Lambda (1 - 3k)}, \quad q = - \frac{3 m (3k + 1)}{2 \Lambda (1 - 3k)}. \]

For $\Lambda < 0$ and $0 < k < 1/3$ the above equation has a real negative root. This is a trivial consequence of the fact that $f(0) = 2q > 0$ and $\lim_{r_\ast \to - \infty} f(r_\ast) = - \infty$. Thus, a real and positive root can exist only if the discriminant $W = q^2 + p^3$ is nonpositive. A straightforward calculation shows that this condition is equivalent to
\begin{equation}
\label{masslimit}
 m^2 \leq \frac{32 k^3}{9 (1 - 3k)(1 + 3k)^2 |\Lambda|},
\end{equation}
which can be interpreted as an upper bound on the black hole mass allowing for the transonic accretion. In this case Eq.\ (\ref{eqf}) has three real roots, two of them being positive. This last statement follows directly from the analysis of the complex roots in the Cardano formula
\[ r_\ast = \sqrt[3]{-q + i \sqrt{|W|}} + \sqrt[3]{-q - i \sqrt{|W|}}. \]

Conversely, for $k > 1/3$ and $\Lambda < 0$ the discriminant $W$ is always positive, and there is just one real root. This root is positive: in this case $f(0) < 0$ and, of course, $\lim_{r_\ast \to \infty} f(r_\ast) = + \infty$.

In principle, the analysis of the critical points can be pursued further by computing the Jacobians of the right-hand side of Eqs.\ (\ref{dyn_a}) and (\ref{dyn_b}), i.e.,
\begin{equation}
\label{jac}
 \left( \begin{array}{cc}
\frac{\partial f_1}{\partial r} & \frac{\partial f_1}{\partial u^r} \\
\frac{\partial f_2}{\partial r} & \frac{\partial f_2}{\partial u^r} \\
\end{array} \right)
\end{equation}
at critical points, and analyzing their eigenvalues. This can be easily done, say with \textit{Wolfram Mathematica}. One can show that the critical points are either a saddle (Jacobian (\ref{jac}) has real eigenvalues with different signs) or a center (eigenvalues of (\ref{jac}) are imaginary). More precisely, it can be shown that the eigenvalues are of the form $\pm \sqrt{Y}$, but the resulting expression for $Y$ is lengthy, and the sign of $Y$ is not immediately clear. A case-by-case study shows that the critical point with smaller $r_\ast$ is a saddle point ($Y > 0$), and the critical point with larger $r_\ast$ is a center ($Y < 0$), as expected. For the data depicted in Fig.\ 1, i.e., $k = 1/4$, $m = 1$, and $\Lambda = -5/100$, the critical points are located at $r_\ast = 4.05608$ and $r_\ast = 8.34795$. In the first case the eigenvalues of (\ref{jac}) are $\pm 0.391511$. In the second they are $\pm 0.68955 i$.

The mass limit (\ref{masslimit}) is illustrated in Fig.\ 2. It shows a sequence of transonic solutions obtained for increasing values of the black hole mass and the equation of state $p = e/4$. The solid line in this graph depicts the limiting case with the mass $m = m_L \equiv \sqrt{32/(441 |\Lambda|)}$. Note that the ``homoclinic loop'' gets smaller and smaller with the increasing mass, and it disappears for  $m = m_L$.

\section{Polytropic solutions}
\label{sec_poly}

Sample numerical solutions with polytropic equations of state were given in \cite{mach_malec_karkowski}, exhibiting characteristic homoclinic behavior. Here I note that no global polytropic solutions exist (i.e., Eqs.~(\ref{www}) cannot be satisfied asymptotically), irrespective of the assumed parameters of the polytropic equation of state. This can be seen immediately, if one attempts to repeat the reasoning of Sec.\ \ref{sec_asymptotic}. From Eqs.\ (\ref{www}) and (\ref{wwz}) one has
\begin{eqnarray*}
h \sqrt{1 - \frac{2m}{r} - \frac{\Lambda}{3}r^2 + (u^r)^2 } & = & \\
\left(1 + \frac{\Gamma K}{\Gamma - 1} \rho^{\Gamma - 1} \right) \sqrt{1 - \frac{2m}{r} - \frac{\Lambda}{3}r^2 + (u^r)^2 } & = & \mathrm{const}. 
\end{eqnarray*}
The square root in the above formula behaves asymptotically at least as $\mathcal O(r)$, due to the $\Lambda$ term. The only way in which this divergent behavior could be cancelled (to yield the constant appearing on the right-hand side) is to have $h$ vanishing sufficiently fast. This is clearly impossible since $h > 1$. Of course, the unity in the expression for $h$ comes from the $\rho$ term in the formula for the energy density $e$, i.e., a contribution due to the rest-mass of the particles of the gas.

\section{Discussion}

\begin{figure}
\begin{center}
\includegraphics[width=\columnwidth]{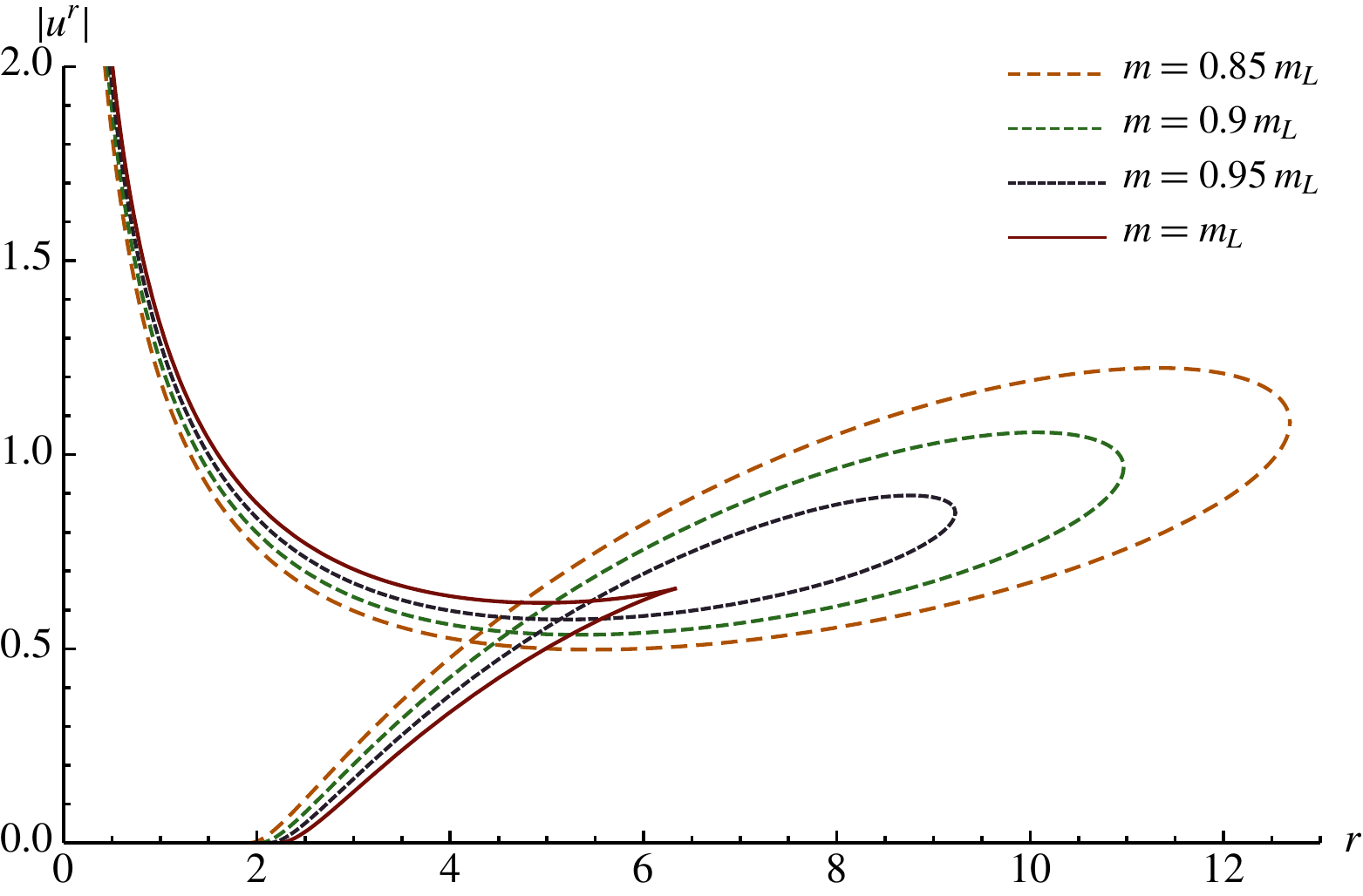}
\end{center}
\caption{Transonic solutions obtained for the equation of state $p = e/4$ and $\Lambda = -5/100$. The graphs correspond to four values of the black hole mass: $m = 0.85 m_L$, $m = 0.9 m_L$, $m = 0.95 m_L$, and $m = m_L$, where $m_L \equiv \sqrt{32/(441 |\Lambda|)}$ is the maximal mass for which a sonic point exists.}
\end{figure}

A very natural interpretation of the above results is to say that the existence of the global (asymptotic) solutions of the Bondi-type accretion in the Schwarzschild--anti-de Sitter spacetime is not directly related to the algebraic form of the equation of state. Global solutions exist for relativistic matter models that can be associated with the gas of massless particles. In this context it is important to note that the equation of state $p = e/3$, a limiting case among isothermal equations of state, is a simple well-known model of the photon gas. This interpretation is also confirmed by the observation that the term that forbids asymptotic solutions in the case of polytropic equations of state is directly connected with the non-vanishing rest-mass of the gas particles.

Of course, the simplest (and perhaps slightly naive) explanation would be to refer to the behavior of radial null and time-like geodesics in the Schwarzschild--anti-de Sitter spacetimes. They are analyzed in detail in \cite{cruz}. Time-like radial geodesics are described by
\[ \left( \frac{dr}{d\tau} \right)^2 = E^2 - \left(1 - \frac{2m}{r} - \frac{\Lambda}{3} r^2 \right), \]
where $\tau$ denotes the proper time, and $E$ is constant (the energy). Here $1 - 2m/r - \Lambda r^2/3$ plays a role of the effective potential. Since it diverges asymptotically as $r^2$, the range of the motion of the particle with finite energy $E$ is limited. In terms of the coordinate time the above equation can be written as
\begin{eqnarray*}
\left( \frac{dr}{dt} \right)^2 & = & \frac{1}{E^2} \left( 1 - \frac{2m}{r} - \frac{\Lambda}{3} r^2 \right)^2 \\
& & \times \left[ E^2 - \left( 1 - \frac{2m}{r} - \frac{\Lambda}{3} r^2 \right) \right].
\end{eqnarray*}
The equation describing null geodesics has the form
\[ \left( \frac{dr}{dt} \right)^2 = \left( 1 - \frac{2m}{r} - \frac{\Lambda}{3} r^2 \right)^2, \]
meaning that a photon can travel to infinity.

It is also interesting to note that a similar homoclinic behavior was observed in the spherically symmetric models of Bondi-type accretion on Reissner--Nordstr\"{o}m black holes \cite{rnaccretion}. In this case it is caused by the electric charge term, and the homoclinic loop is located inside the horizon of the black hole.

As a by-product of the analysis presented in this note I obtained a restriction on the maximum mass of the black hole that allows for the transonic accretion of isothermal fluids with $0 < k < 1/3$. This bound is given by a remarkably simple formula, and it is discussed in Sec.\ \ref{criticalpoints}.

\section*{Acknowledgements}

I am grateful to Jerzy Knopik and Filip Ficek for many fruitful discussions.

This work has been partially supported by the Polish Ministry of Science and Higher Education grant IP2012~000172 and the NCN grant DEC-2012/06/A/ST2/00397.

\end{document}